\documentstyle{article}
\newcommand{\blankline}{\vskip .3cm}
\newcommand{\f}{\begin{equation}}
\newcommand{\ff}{\end{equation}}

\newcommand{\Et}{ \tilde{E}^{ai}}

\begin{document}
\vfill
\centerline{\LARGE What can we learn from the study of  }
\centerline{\LARGE non-perturbative quantum general relativity? }
\blankline
\rm
\vskip1cm
\centerline{Lee Smolin}
\blankline
 \centerline{\it   Department of Physics, Syracuse University,
 Syracuse, New York  U.S.A.}
 \vskip 1cm
\centerline{\today}

\vskip 2cm

\centerline{Abstract}
\noindent

I attempt to answer the question of the title by giving an annotated
list of the major results achieved, over the last six years, in the
program to construct quantum general relativity using the Ashtekar
variables and the loop representation.  A summary of the key open
problems is also included.

\vfill\eject

\tableofcontents
\vfill\eject

\vfill\eject
\section{Introduction}

Most of what we know about nature at the present time is contained
within the realms of
either quantum field theory or general relativity.  Each of these is
a beautiful,
powerful  and profound theory.   However neither can, because of the
existence of the other, be
said to constitute the basis for a general theory of physics.  Thus,
while Newtonian physics has been overthrown, it has not been replaced;
and it cannot be until we can  invent a synthesis of these two great
theoretical
edifices that can serve as a single foundation for our understanding
of all of nature.  To do this is the problem of quantum gravity.
As such, the problem of quantum gravity is very different from other
problems in which we seek to apply a well defined
theoretical structure to a new phenomena.  It
is more open, and more difficult.

Many people who work on this problem complain about its difficulty,
and about its distance from experiment.  However, I think both
complaints are based on a misunderstanding of the nature of the
problem.   After all, the last time the progress of science
required
a transformation of this scale
in our basic understanding of nature, it took more than 140 years
from the
publication of Copernicus's {\it Revolutionibus} to the publication of
Newton's {\it Principia} \cite{Julian-book}.
could 
As far as experiment is concerned, there is a large amount of
data about fundamental physics that is presently unexplained,
including  the undetermined parameters of the
standard model of particle physics, the horizon and flatness
problem of cosmology and  the problem of explaining the
formation of structure in the universe.  There are good
reasons to believe that a quantum theory of gravity will
have something
important to contribute to the solution of each of these problems.

Many different approaches to this problem have been pursued.
This  is proper, as
there is no way of really knowing from what direction the
solution will come.  Moreover,
as the philosopher of science Paul Feyerabend reminds us,
science functions best
when the level of consensus remains near the minimum forced on us by
the experimental data\cite{Feyerabend}.   In this contribution,
I want to discuss only one  approach to the problem.
This approach has been particularly active during the
last six years and it has
achieved a certain amount of progress, which I want to
summarize here.  There
are also big unsolved problems, which I will also be mentioning.

The approach I want to describe can be characterized by a
list of questions
that those who pursue it are seeking to answer.  These
may be stated as follows:

1)  We take it as given that any quantum theory of gravity that has a
chance of being a correct description of reality must be non-
perturbative.    This means that the theory cannot be based on an
expansion around a single classical background, in which it is the
deviations from the background that are quantized.  Instead, all of
the geometrical quantities that describe the geometry of spacetime
must be treated as quantum operators.   This means that very few
of the techniques of
conventional quantum field theory can be directly applied to it.
Can we invent a
new approach to quantum field theory that applies to the cases
of field theories
defined on differential manifolds without fixed metric structures?

2) Specifically, it is  diffeomorphism invariance that expresses
the absence of a non-dynamical
background geometry in classical general relativity.  As
such, its implementation in the quantum theory is the central
problem of constructing any quantum theory of gravity.
 We would then like to understand how the requirement of
exact diffeomorphism invariance requires us to modify the standard
quantum
field theory techniques such as regularization, renormalization,
operator
product expansions, and so forth.

3)  We are not trying to construct {\it the} quantum theory of gravity.
We are trying to construct {\it a} theory which is consistent with
quantum
mechanics and general relativity.  For this reason we study quantum
general relativity because  we have its exact nonperturbative
formulation, classically.  String theory is, in principle, an
attractive
program for the unification of physics.  However, as long as it lacks
a nonperturbative formulation it cannot be the basis of an
exploration of the problems of nonperturbative approaches to
quantum gravity.
We will not be disappointed if, in the end, quantum general
relativity is not the theory of nature.  But we aim to decide cleanly
whether or not there is a consistent mathematical theory, with a
sensible physical interpretation, that could be given that name.

4)  Certainly it may be true that a complete solution to the
problems of
quantum gravity and quantum cosmology will require also a solution
to the problem of the unification of gravitation with the other
fundamental interactions.    This could happen through a "microphysics
$\rightarrow$
macrophysics" approach like string theory or a "macrophysics
$\rightarrow$ microphysics"  approach, one example of which is
proposed in \cite{re-evolution}.
At the same time, it may also be the case that important aspects of
the problem of quantum gravity can be solved independently of
what matter fields
gravitation is coupled to.  This will be the case to the extent that
qualitatively new physics emerges from the
construction of diffeomorphism invariant quantum field theories.
We would like to
see if the study of quantum general relativity at the nonpertubative
level can
lead to the discovery of such phenomena.

5)  In any such approach, one of the key questions is what
constitutes  an
observable.  This is because, in a diffeomorphism invariant theory,
coordinates
and clocks have no {\it a-priori} meaning; any meaningful observable
must express some
relationship between physical fields, rather than being defined
with respect to a
background geometry or an external observer.  One aspect of this
problem is the
problem of time in quantum cosmology\footnote{Good discussion of
this problem
are in \cite{karel-review,Carlo-time}.
The author's point of view about the
problem of time in quantum cosmology
is described in \cite{time-evolution}.}.
These problems arise already at the classical
level,
but there they can be
solved\cite{Carlo-time,Bryce-matter,carlo-matter,karel-matter,
time-evolution}.  We then need to ask:  Can we put what we
know about the problem of observables and time in classical general
relativity together with
the technical developments in diffeomorphism invariant quantum field
theory to learn how to
construct operators that correspond to physically meaningful
quantities in a quantum theory of gravity and cosmology?

While these are the main questions that have guided us in the work
I will be describing below, the fact that we have been able to make
any progress is
 due primarily to two technical developments.  These are the
 Ashtekar formulation
of general relativity \cite{Abhay}, and the loop representation of
quantum field
theories \cite{carlolee,gambini-loop}.

The Ashtekar variables and the loop representation have been the
subject of a number of reviews, including three published within the
last two years\cite{review-abhay,review-carlo,review-lee}.   For
this reason, I will not repeat here what the reader can
easily find in those reviews or in the original papers.  Instead,
I  will
devote myself
to describing, in as concise a manner as possible, the basic
results that have been, so far, achieved  by this program.  As
my aim is to be brief, I do not give
many technical details, except when describing results that have
not so
far appeared elsewhere.

I should point out that while I have tried to write this paper so
it can be read by someone who is not an expert on quantum gravity,
I do not include an introduction to
the basics of
canonical general relativity and the methods of canonical
quantization. These can be found in other papers in this
proceedings, such as the paper of Kuchar\cite{karel-this}, as
well as in numerous
other places, among
them\cite{review-abhay,review-carlo,review-lee}.

I begin in the next section by asking how quantum field theory must
be modified so
that it can make sense in the absence of a background metric.  The
core of the paper
is in the following two sections, where I give, correspondingly, an
annotated  list of major results
and a list of key open problems.  In the conclusion I try to provide
an answer
to the question in the title\footnote{I want to mention that this is
not intended as a review of all of the developments associated with
the Ashtekar variables.  Particularly, there are several important
developments connected with the classical
theory that most likely have implications for the quantum theory,
such as the Capovilla-Dell-Jacobson\cite{CDJ} formalism and the
classical solutions of the diffeomorphism constraints of Newman and
Rovelli\cite{tedcarlo}.  However, as my focus is the quantum theory,
and as their implications for that have yet to be developed I do not
mention them, as well as a number of other very interesting
developments,  here.}.

\section{Basic ideas of the approach}

A good place to begin is with the question of how to construct
diffeomorphism invariant quantum field theories.  The key problem
is how to get rid of the dependence on the background metric that
underlie the standard formulations of quantum
field theory in Minkowski spacetime.  Here are four ways in which
Minkowski spacetime quantum field theories depend on the
background metric:

{\bf A)}  The Fock spaces of linearized field theories are
constructed by
associating operators with the solutions of free field theories
on a
background spacetime.

{\bf B)}  The definition of the vacuum and the definition of the
creation
and annihilation operators depend on the splitting of the solutions of
the linearized field equations into positive and negative frequency
parts.  This splitting is Poincare invariant but, as we know from our
experience with quantum field theories in curved spacetime, it is not
invariant under any larger group of transformations and depends
also on the background metric.

{\bf C)}  The regularization and renormalization procedures necessary to
make sense out of operator products and interactions in quantum
field theory all depend explicitly on the background metric.  The
background metric is used in the definition of the cutoff scale, in the
separation of terms in the operator product expansion and to
measure how fast the cutoffs are removed in taking the limits that
define renormalized operator products.

{\bf D)}  The inner products of conventional quantum field theories are
defined by the requirement of Poincare invariance with respect to a
given flat background metric.

We propose to replace each of these constructions with alternatives
that do not depend on any background metric.   The way in which we
go about doing it defines the approach we are calling non-
perturbative quantum general relativity.  These alternative
constructions are:

{\bf A')} We construct background independent quantum field theories by
constructing new representations of algebras of observables that are
unitarily inequivalent to the Fock representations.  It is here that the
loop representation plays the key role.  These representations carry
unbroken unitary representations of the diffeomorphism group; this
makes possible the exact solutions of the constraints that impose
diffeomophism invariance on the quantum states.

{\bf B')}  We replace the splitting into positive and negative frequency
parts by a splitting into self-dual and antiself- dual parts.
This latter
splitting is defined only in special cases such as that of connection
fields in four spacetime dimensions.  It has the great virtue that the
connection and curvature of any four dimensional spacetime can be
split into self-dual and antiself-dual parts; as a result, the self-dual
representation, in which states are functions of the self-dual part of
the field, can be defined in a nonperturbative quantization.  This
stands in contrast to a positive frequency representation, which can
only be defined with respect to a fixed background metric and a
correspondingly restricted time coordinate.

This is one of the main motivations for the use of the Ashtekar
variables as the basis for the quantization.

At the linearized level, the self-dual part of a connection (whether or
the electromagnetic, Yang-Mills or linearized gravitational field)
consists of the positive frequency part of the left handed helicity,
plus the negative frequency part of the right handed helicity.   The
negative frequency part must then be quantized in a kind of "anti-
Bargmann" representation.  It was not entirely obvious that such a
quantization exists, and one of the key results of the program is that
self-dual representations exist at the linearized
level\cite{sd-us,gravitons}.

{\bf C')}  As I will describe below, the process of renormalization,
through
which the product of local operators is defined to be another local
operator, is necessarily dependent on a background metric (or at
least on a background volume element \cite{review-lee,weave}.)
There is then no local
background independent renormalization procedure for local
operators.  At the same time, we have found that there are
background independent regularization procedures for certain non-
local observables.  Their use results in finite and background
independent
operators\cite{weave,review-lee}.  Further, one can make an argument
that any operator constructed from functions of local operators
through a regularization procedure which is diffeomorphism
invariant in the limit that the regulator is removed will necessarily
be finite\cite{review-lee,carloun}.

{\bf D')}  Without a background metric, there is no symmetry principle
that can guide the selection of the inner product.  We propose to base
the selection on an alternative principle:  a complete set of
real physical observables must be represented by self-adjoint
operators\cite{review-abhay,review-carlo}.
The proposal depends on the construction of  a
set of physical observables, realized as well defined operators on the
space of physical states, for which the classically corresponding
observables are known.  We can then use the reality conditions
satisfied by the corresponding classical observables to posit
hermiticity relations for the physical operators.  The inner product is
then to be determined by the condition that it realize these
hermiticity conditions.

Clearly, what I have just stated is a strategy, but it is not a
completely defined procedure.  Crucial questions such as which
observables and how many observables are necessary are left
unspecified.  In a number of special cases, including Maxwell
theory\cite{abhaycarlo-maxwell,sd-us}
and linearized gravity\cite{gravitons}, in 2+1
gravity\cite{2+1-us,2+1-me,review-abhay}
 and other finite
dimensional examples \cite{tate-constrained}, this principle can
be implemented and leads
to the physically correct inner products.  However, for reasons that will
become clear later, we have not yet been able to
test this idea in the case of full quantum gravity.

\section{The main results of nonperturbative quantum gravity}

I would like now to state, concisely and with a minimum of
technicalities, exactly what the main results of this approach are,
to date.  To frame the discussion, let me begin by recalling the main
elements of canonical quantization of general relativity in the Ashtekar
formalism,
following the method of Dirac\cite{dirac}.  We begin with a phase space,
which is
taken to be the space of pairs of complex $SU(2)$ connections,
$A_a^i$ and conjugate electric fields $\tilde{E}^{ai}$ on a three
manifold $\Sigma$ of fixed topology.  For everything that follows it
will be crucial that $\Et$ is a vector density field.  This is necessary
so that the Poisson bracket relation,
\f
\{A_b^j (y) , \Et (x) \}=\delta_b^a \delta^{ij}\delta^3 (x,y)
\ff
makes sense, because the delta function  must, in the absence of a
background metric, be a density.

To quantize the system, we proceed to construct the state space in
three steps, which we call kinematic, diffeomorphism invariant and
physical.   To
define the kinematical state space, ${\cal V}_{kin}$, which is the
starting point for the
quantization, one must first pick out of the algebra of functions on
the phase space a subalgebra that one wants to have represented
exactly by quantum operators.  A choice is necessarily involved at
this stage; because of the existence of problems of ordering and the
regularization of operator products, it is impossible that the whole
algebra of functions on the phase space be identically represented as
an algebra of operators.  We shall call the subalgebra chosen $\cal
A$.  The one condition it should satisfy is that its elements should
coordinatize the phase space of interest.

Once we choose $\cal A$, the kinematical state space is then
constructed by finding a linear space ${\cal V}_{kin}$ on which
there acts an
algebra $\hat{\cal A}$,
that is isomorphic to $\cal A$, at least up to terms that vanish in
the limit that $\hbar \rightarrow 0$.

The diffeomorphism invariant and physical state spaces, which will be
called
${\cal V}_{diffeo}$ and ${\cal V}_{phys}$ are then constructed as
follows.  One constructs operators in ${\cal V}_{kin}$  that correspond
to
the classical diffeomorphism and Hamiltonian constraints.  (These
will be denoted
$\hat D(v)$
and $\hat H(N)$, respectively, where $v^a$ and $N$ are smearing
functions that are, respectively, a vector field and an inverse density
field.)
${\cal S}_{diffeo}$ is then defined to be the subspace of states in
${\cal V}_{kin}$ that are annihilated by $\hat D (v)$.
${\cal V}_{phys}$ is
defined to
be the subspace of those states that are annihilated both by $\hat D
(v)$ and
$\hat H (N)$.

This process is called Dirac quantization\cite{dirac}.  Unfortunately
Dirac, while
setting
out the procedure in one of his very readable little books, failed to
include
two details that must arise in any field theoretic application of the
method.
The first is that, at least in all representations known to this time,
 the Hamiltonian constraint involves operator products and must be
regularized.  The second is the question of how to choose the inner
product.

The key point is that it is not sufficient to give an inner product at
the
kinematical level, because, unless one is very lucky, the solutions to
the
quantum constraint equations will be non-normalizable with respect
to the
kinematical inner product.  (Thus, ${\cal V}_{diffeo}$ and
${\cal V}_{phys}$ are, in general, subspaces of ${\cal V}_{kin}$
as vector spaces, and not as Hilbert spaces.)  The physical inner
product must be
chosen just
on the space ${\cal V}_{phys}$ (and similarly for the diffeomorphism
invariant states).

Let us now begin the quantization along these lines.
Conventionally, one choices to quantize using the algebra (1).  This is
the starting point of the construction of Fock representations.
However, to construct the non-Fock representations we will be
interested in, we take a different subalgebra as a starting point.  The
main idea behind this choice is to implement the $SU(2)$ gauge
invariance explicitly by choosing an algebra of gauge invariant
functions.  To do this we take  the configuration variable is taken to
be the holonomies,
\f
T[\gamma ] \equiv {1 \over 2} Tr P e^{G\int_\gamma A}.
\ff
For a conjugate variables we would like to take a family of functions
linear in the conjugate momentum.    The requirement of gauge
invariance then requires us to take functions that are also dependent
on loops.  To construct them, consider a loop $\beta$ on which we
have a preferred point $\beta (s)$.  We may then define the function,
\f
T^a [\beta ](s) \equiv {1 \over 2} Tr \left [
\tilde{E}^a \left (\beta (s) \right )
Pe^{G \int_s dt \dot{\beta}^a(t) A_a(t) }  \right ]
\ff
defined by tracing $\tilde{E}$ at the point $\beta (s)$ against the
holonomy around $\beta$ starting and ending at $\beta (s)$.  To
complete the definition of the conjugate variable, let us consider a
rubber band,
$\bf \beta$ defined by  one parameter family of loops $\beta_u (s)$
with
$0 \geq u \geq 1$.  We may then define
\f
E[{\bf \beta }] \equiv \int \int du ds \epsilon_{abc}{ d\beta_u^a(s)
\over ds}
{d\beta_u^b  (s)\over du} T^c [\beta_u ](s)
\ff

The two observables (2) and (4) have a very pretty algebra.  Let us
define
\f
 I[\gamma , {\bf \beta }]= \int d\gamma^a (t)
\int d^2 {\bf \beta }^{bc}(s,u)
\delta^3 (\gamma (t) , {\bf \beta } (s,u) ) \epsilon_{abc}
\ff
to be the intersection number of the loop $\gamma$ with the two
dimensional surface $\bf \beta $.  The intersection number is equal
to
an integer; it counts, with the sign reflecting the orientation, the
number of
times that the loop intersects the strip.   A simple calculation than
yields,
\f
\{   T[\gamma ], E[{\bf \beta }] \} = G  I[\gamma , {\bf \beta }]
\left [  T[\gamma \circ \beta_{u^*} ]  - T[\gamma \circ \beta_{u^*}^{-
1} ] .  \right ].
\ff
Here, $\alpha \circ \beta $ refers to the loop made by combining
$\alpha$ and $\beta $ and $\beta_{u^*}$ is the element of the one
parameter family that goes through the point of the rubber band
where $\gamma$ intersects it.

This algebra is called the loop-strip algebra, or the loop algebra for
short.

It is important to comment on the presence of Newton's constant,
$G$, in the
definitions of the holonomies in these observables.  Because the
frame field $\Et$ should be dimensionless, the quantity it is
conjugate to, the Ashtekar connection,  cannot have the usual
dimensions of a connection of inverse length if the Poisson brackets
(1) are to hold.  Instead, it is $G A_a^i$ that has dimensions of
inverse length.  As a result, there is a $G$ in (6).  We will see
later that this plays a key role in
several results described below.

With this background, I now give an annotated list of the major
results so far
which have been achieved in the direction of nonperturbative
quantum general relativity.  The results will be organized according
to the three levels of Dirac quantization:  kinematical,
diffeomorphism invariant and physical.

\subsection{Existence of the loop representations at the kinematical
level}

A loop representation is a quantization of a gauge theory based on
the loop
algebra given by (6).  The loop algebra bears a relationship to the
canonical algebra (1) which is somewhat analogous to that the Weyl
algebra  bears the simple Heisenberg algebra.  There are
representations of the loop algebra which are also representations of
the canonical algebra.  Among them is the Fock representation.
However, there are also representations of the loop
algebra which in which there exist no operator linear in $A_a^i$,
which are inequivalent to the Fock representation.  Among these are
the representations which are of interest to non-perturbative quantum
gravity\cite{rayner,abhaychris}.

A loop representation may be characterized as follows.  We introduce
a
basis of bra states, labeled by loops, such that any ket state $|\Psi >$
may be
written as
\f
\Psi [\gamma ] = <\gamma | \Psi >
\ff
Note that this notation does not assume the existence of an inner
product.
The elements of the ket space are functions $\Psi [\gamma ]$ which
live in
some space of functions, the complete specification of which is
necessary to
define the representation.  The bra states, $< \gamma |$ are a linear
space
dual to the ket space, whose action is defined by (7).  For any loop
representation these ket states are constrained to satisfy
\f
\sum_i c_i <\gamma_i | =0
\ff
whenever the holonomies satisfy
\f
\sum_i c_i
TrPe^{\int_{\gamma_i}A} = 0   \nonumber
\ff
for all connections $A_a^i$.  This is the way that the identities
satisfied by
holonomies-the Mandelstam identities-are imposed on the
representation.
The loops $\gamma$ may then be taken to be piecewise
differentiable loops.  It is convenient also to use a convention in
which a loop can refer to a set of
loops, if we assume that the trace of the holonomy of a set of loops is
taken
to be the product of the traces of the holonomies of the individual
loops.
Indeed, by a loop or a set of loop we really always mean equivalence
classes
of loops under the relation (8)\footnote{As has been emphasized by
Gambini and Trias and their coworkers, the space of these
equivalence classes of
loops actually forms a group\cite{gambini-loop}.  The basis for
their alternative
formulation
of the loop representation is the representation theory of this
group.}.

The action of the operators that represent the loop variables (2) and
(4) are
then defined by their action on the bra states:
\f
<\gamma | \hat T [\alpha ]= <\alpha \cup \gamma |
\ff
\f
<\gamma |  \hat E [{\bf \beta }]= \hbar G \left (
<\gamma \circ \beta_{u^*} | - <\gamma \circ \beta_{u}^{-1} |\right  )
\ff
In these equations the $\cup$
represents union in a set of loops while
the
$\circ$ represents forming a new loop by a product of two loops.  It
is not
hard to verify that these operators satisfy the algebra (6) (with the
$\hbar$
inserted.  It is, in fact, interesting to note that $l_p^2=\hbar G$
appears
in the algebra even at the kinematical level.  This fact (and the fact
that
it is the Planck area and not the Planck length that appears) plays a
crucial role below.

Finally, we may
note that the fact that
the group is  $SL(2,C)$ (or some real subgroup of it) appears in the
form
of the right hand side of (6), together with the equivalence relations
generated by (8).
Connections based on other Lie groups also have loop
representations; they
differ only in the form of this action and in the equivalence relations
generated by the holonomies.

\subsection{Applications of the loop representations to linearized
field theories:  Fock representations}

An example of the forgoing is the loop representations of abelian
connections.   The loop representation actually first appeared in a
quantization of the free Maxwell field given by Gambini and Trias in
1981 \cite{gambini-loop}.  The loop quantization of free Maxwell
theory was also treated
later in \cite{abhaycarlo-maxwell,sd-us}.
In these papers it is shown that one can construct a space
of states $\Psi [\gamma ]$ which is isomorphic to the Fock space of
free photon states.  The whole of free quantum electrodynamics,
including
the Hamiltonian, creation and annihilation operators and inner
product
may be written simply in the loop representation.

Indeed, there are several different representations of the Fock space
as functions of loops.  These correspond to the different ways to
write the Fock space as functions of the connection.  The two well
known connection representations of free field theories are the
Schroedinger representation, in which the states are functions of the
real connection, $A_a$ and the positive frequency, or Bargmann
representation, in which the states are functions of the positive
frequency part of the connection.

In the loop representation that corresponds to the positive frequency
connections the ground state is simply written as
\f
\Psi_0 [\gamma ] = <\gamma |0> = 1
\ff
The state of one photon of momentum $\vec p$ and polarization
$\vec \epsilon$ is written,
\f
\Psi_{\vec{p}, \vec{\epsilon}}[\gamma ] = <\gamma |\vec{p},
\vec{\epsilon}>
=\oint ds e^{ p_b \cdot \gamma^b (s)} \ \dot{\gamma}^a (s) \epsilon_a
\equiv F[\vec{p}, \vec{\epsilon}]
\ff
The $  F[\vec{p}, \vec{\epsilon}]$'s play a dual role in the formalism.
First, they provide a useful set of coordinates for the loops modulo
the relations (8), with the abelian holonomy\footnote{Their extension
to a set of coordinates
on the space of loops modulo the relations that arise from non-
abelian holonomies are the basis of a new approach to the loop
representation of Gambini and his collaborators\cite{gambini-newloop}.}
Second, the
multiphoton states are written as polynomials of the $F[\vec{p},
\vec{\epsilon}]$'s.  The Fock space is then defined to be the space of
functions of loops that are analytic functions of the loop coordinates
$ F[\vec{p}, \vec{\epsilon}]$.

The Hamiltonian and inner product of quantum Maxwell theory
are then written as operators on loop space as follows,
\f
\hat H = \sum_{\epsilon} \int d^3p|p|F[\vec{p}, \vec{\epsilon}]
{\delta \over { F[\vec{p}, \vec{\epsilon}]}}
\ff
\f
<\Phi |\chi > = \int [dF]\bar{\Phi}\chi
e^{-\int {d^3p\over |p|}\left | F[\vec{p}, \vec{\epsilon}] \right |^2}
\ff

It is interesting to note that in every case where the construction of
the loop
representation has been completed, it has been found that there
exists a transform that connects it to the appropriate connection
representation in which the states are functions of the connection.
The general form for this
transform is
\f
\Psi [\gamma ] = \int d\mu [A] T[\gamma ,A] \psi [A]
\ff
where $d\mu [A]$ is an appropriate measure on the space of
connections mod gauge transformations.

\subsection{Existence of self-dual representations for both
connection and loop representations}

The key discovery of Ashtekar is that in full general relativity the
self-dual part of the connection may be considered as a configuration
variable of the theory, i.e. every component at ever point commutes
with every other one.  Because of this, we would like to base  the
quantization of the full theory on  the self-dual representation, in
which the observables, (2) and (4), whose algebra we quantize, are
taken to be functions  of the self-dual part of the connection.
If this is to be a successful route to the quantum theory it would be
most convenient if the linearized theory can also be quantized in a
representation in which the observables are functions of the self-dual
part
of the linearized connection.  This, however, gives rise to the
following question:  At the linearized level, the self-dual part of the
connection is the positive frequency part of the left handed helicity
component, plus the negative frequency part of the right handed
helicity component.  This means that the right handed component
must be quantized in a kind of negative frequency, or anti-Bargmann
representation.   Thus, the first question that must be asked is
whether there exist such anti-Bargmann representations, which are
diagonal in the annihilation operator rather than in the creation
operator.  At first sight this seems to be impossible; consider for
example the equation that the annihilation operator annihilates the
ground state.  In such a representation it must read
\f
<\bar z | \hat a |0> = \bar z \psi_0 (\bar z ) =0.
\ff
In fact, such representations exist, but their expression requires
distributions rather than holomorphic functions, as in the usual
Bargmann representation\cite{sd-us}.  Thus, in the sense of
distributions, the
solution to (17) is
\f
\psi_0 (\bar z ) = \delta (\bar z ).
\ff

Once this obstacle is overcome, it is straightforward to construct the
full
negative frequency representation, for the harmonic oscillator and
for any linear field theory\cite{sd-us} whose configuration variable
is a connection.  Among these is linearized
gravity\cite{review-abhay}, which
may be quantized in the loop representation\cite{gravitons}.

The existence of the loop representation for linearized general
relativity not only serves as a confirmation of the basic program of
nonperturbative quantum gravity, it may expected to play a key role in
the physical interpretation of the exact theory.

\subsection{Non-Fock representations of the loop algebra}

The Fock representations are important for the construction of
linearized field theories, but they are inappropriate as a starting
point for the construction of diffeomorphism invariant theories.  This
is because they cannot carry unbroken representations of the
diffeomorphism group for the simple reason that the
diffeomorphisms are broken by the existence of the background
metric.  Since the diffeomorphisms cannot be represented one cannot
use them as a starting point to construct diffeomorphism invariant
states.

The key question is then, do there exist representations of the
kinematical observable algebra that carry unbroken representations
of the spatial diffeomorphism group?  I do not know the answer for
the case of the canonical algebra (1).  But, for the loop algebras of the
form of (6) (and their generalizations to other groups) there do exist
representations with this property\cite{rayner,abhaychris}.

These representations are based on the use of the discrete measure
on the space of loops\footnote{recall that by loops I always mean
loops modulo the relations (8).}

To construct this representation it will be useful to introduce a set of
basis states, which we call characteristic states.  For a loop $\alpha$
which contains no intersections, we may define the state, such that,
for $\gamma$ also nonintersecting,
\f
\chi_\alpha [\gamma ]= <\gamma |\alpha >= 1
\ \ \ {\rm if} \ \ \alpha =
\gamma \rm{and \ \ otherwise \ \ vanishes}.
\ff
Here, the equality is always meant in the sense of the equivalence
relations (8).   There is one more case, which is  if $\gamma \neq
\alpha$, but contains intersections.  In this case $\chi_\alpha
[\gamma ]$ does not necessarily vanish, its actual value is
determined by requiring that it be an eigenstate of the operators defined
in (25) and (30), below \cite{review-lee}.  There are also basis states
associated with intersecting loops \cite{review-lee}.

We will denote these characteristic states abstractly by
$|\alpha >$, making use of the standard Dirac formalism in which
bras represent elements of function spaces and kets are linear maps
from those function
spaces to the complex numbers.

Let us then consider the linear space, ${\cal V}_{discrete}$, which
consists of states of the form,
\f
\Psi [\gamma ] = \sum_I c_I \chi_{\alpha_I}[\gamma ]
\ff
where we require that
\f
\sum_I |c_I|^2 < \infty
\ff
Here the sum is over any countable set of loops  in $\Sigma$.
It can be
easily verified that the formal definitions of the loop operators,
(10) and (11), are well defined when acting on the states in
${\cal V}_{discrete}$.

We can impose an inner product on ${\cal V}_{discrete}$ by defining
\f
<\alpha |^\dagger = |\alpha >
\ff
Note that this inner product does not realize the kinematical relativity
conditions for the $T[\gamma ]$ observable\footnote{For the reader
unfamiliar with the reality conditions, they are the conditions that
the three metric and its time derivative both turn out to be real
when computed in the Ashtekar formalism.  They imply
that
$A_a^i$ is a complex connection, which, together with its complex
conjugate, satisfies a certain polynomial condition.  The result is
that the loop operators are not real.}, but it does realize them
(at least formally) for functions of $\tilde{E}^{ai}$ only.  Like any
kinematical inner product, it is useful only
as a mathematical device.

With any choice of inner product which makes the characteristic states
normalizable, the discrete representation is unitarily inequivalent to
Fock space.   There is a possibility that it may be of use for
nonperturbative treatments of gauge theories, because it
implements, in the continuum, the quantization of the electric flux.
This is a possibility that needs further development.  At the present
time its use comes from its application to diffeomorphism invariant
quantum field theories.  This is because an exact, unbroken, unitary
representation of the diffeomorphism group can be defined on it as,
\f
\hat U (\phi ) |\gamma > = |\phi^{-1} \circ \gamma >
\ff
where $\phi$ is any diffeomorphism.
This means that the generator of diffeomorphisms is well defined in
this representation,
\f
\hat D(v) \Psi [\gamma ] = {d \over dt} \hat U (\phi_t) \Psi [\gamma
],
\ff
where $\phi_t$ is a one parameter group of diffeomorphisms
generated by the vector field $v^a$.  $\hat{D}(v)$ may be shown
from these definitions to
satisfy the algebra of vector fields on $\Sigma$.

\subsection{Classification of the loop representations in the real case}

I have described two different representations of the loop algebra (6).
In one case, in which we require that both both $A_a^i$
and
$\tilde{E}^{ai}$ are real, the representations of the $SU(2)$
loop algebra
have
been completely classified.  This was done by Ashtekar and Isham
 \cite{abhaychris}, who make use of the fact that in  this case the
 loop algebra (6) is a star
algebra.  The classification can then be
done using some of the technology of the
representation theory of star algebras developed by Gel'fand and
collaborators.

\subsection{Application of the loop representation to quantum Yang-
Mills
theory}

As I mentioned above, the loop representation may be applied to
non-abelian gauge theories\cite{latticeloops}.  The loop
representation may be
developed in the context of the lattice regularization, where the loop
states provide a
gauge invariant basis for the state space.  This formulation has been
the
starting point for several works in which new numerical approaches
to
lattice gauge theory, both with and without fermions have been
explored.
These works involve approximation procedures which are based on
the fact
that in the loop basis almost all the matrix elements of both the
Hamiltonian and inner product are
zero .  As a result,  sparse matrix and cluster techniques may
be
applied\cite{bernd-lattice,Farrerons}.
For example, in $2+1$ dimensions extensive numerical calculations
have
been done for both $SU(2)$ \cite{bernd-lattice} and
$SU(3)$ \cite{Farrerons},
which showed that
results for the ground state energy and mass gap (as functions
of the coupling constant), obtained previously by Monte Carlo
simulations
and  are reproduced accurately.  Furthermore, in $3+1$ dimensions
numerical work has been, and is being, done for the case of $QED$
with
fermions\cite{gambini3+1fermions}.

In addition to these numerical approaches, there have been some
very
interesting analytical work done on non-abelian gauge theories in
the loop representation, by Loll\cite{renata}, Rovelli\cite{carloun}
and others.

\subsection{Nonexistence of local operators in non-perturbative
quantum
gravity}

For the remainder of this section, I will confine myself to the
applications of
the loop representation to quantum gravity.  I begin with several
results about observables and the classical limit at the kinematical
level.  First,
of all, it is not trivial to construct quantum
observables at the kinematical level because such observables must
be invariant under the Yang-Mills gauge transformations, and any
such observables that involve the frame fields involve operator
products.  Thus,
regularization is an issue even at the kinematical level.

As the kinematical level is meant to be a stepping stone to the
diffeomorphism invariant and physical levels, we will be interested
only in regularization procedures that do not introduce extra
background
dependence into the final definitions of the operators.  Any
regularization procedure depends on additional structure such as
background metrics or coordinate systems as these  are needed
to specify how the point splitting is done or define the cutoffs.
What we must then require is
that
when finite operators are finally produced as a result of the process
they
have no dependence on these structures.

We have discovered that this requirement seems to rule out the
conventional renormalization procedures of Poincare invariant
quantum
field theories\cite{weave,review-lee}.  Although this was
discovered through a painful
process
in which many possible approaches were tried and discarded, the
reason
for this can be stated very simply.  Local
operators are distribution valued and
distributions are, in the absence of a background metric, densities of
weight one.  A renormalization procedure is a procedure by which a
product of two local operators is defined to be a third local operator.
It is thus a procedure for multiplying two distributions to get a third
distribution.  However, there is a problem with the density weights,
because the product of two distributions should have density weight
two, but a local operator will have only density weight one.  The
result
is that any such renormalization procedure must introduce an
additional
scalar density so that the density weights on the left and right hand
sides
of the product match.  What we found was that in any procedure we
tried,
such a density always appeared which was a function of the
background
structures introduced in the regularization and renormalization
procedures.

Now, in Minkowski spacetime, or even in quantum field theory in a
curved spacetime,  there is a preferred density which is given by the
determinant of the background metric.  In these cases the ambiguity
may be reduced to one free renormalization constant. This is the
reason for the existence of a free renormalization scale in the
conventional renormalizations of operator products.

However, in nonperturbative quantum gravity there is no preferred
density and the ambiguity of a scale in the renormalization of a
quantum field theory in a classical background becomes an
ambiguity up to a density.  The result is that it is very difficult to
imagine how a renormalization procedure could be constructed for
operator products in this context that did not lead to a breaking of
diffeomorphism invariance.

This means, in particular, that when using a frame field
formalism such as the Ashtekar variables there is no operator
to measure the metric at a point.  This is because the basic
variable is the frame field, $\tilde{E}^{ai}$, which is related to the
metric through
$\tilde{\tilde{q}}^{ab}=\tilde{E}^{ai}\tilde{E}^b_i$.

\subsection{Existence of finite, background independent non-local
operators at the kinematical level}

One might think that as a result of the situation I've just described it
is impossible to define meaningful observables that measure the
spatial metric in non-perturbative quantum gravity.  Fortunately,
this is not the case, because there are non-local observables that are
equivalent to the metric in the sense that a complete measurement
of them allows the metric to be reconstructed.   We have found that
it is possible to construct quantum operators that correspond to some
of these non-local observables and that these operators are finite and
background independent, when constructed by means of the right
regularization procedure\cite{weave,review-lee}.
Thus, as these operators don't need to be
renormalized, they escape the difficulty I described in the previous
paragraph.

The idea behind the construction of these operators is very simple:
If there is no unambiguous procedure for multiplying two
distributions to get a third distribution, we may construct
unambiguous procedures that define the {\it square root of the
product of two distributions.}

I will mention here three examples of such
observables\cite{weave,review-lee}.
First, given
any one form $\omega$, we may define the integral of its norm as
follows,
\f
Q[\omega , \tilde{E}^{ai} ] = \int_\Sigma
\sqrt{\tilde{E}^{ai}\omega_a\tilde{E}^{bi}\omega_b}
\ff

This observable can be regulated through a modified point splitting
procedure.  I will not describe it here, the details are given
in \cite{review-lee}.
As may be expected, the hardest part of the construction is taking
the operator square root.   The result is easiest to express in terms of
the bras $<\alpha |$.  For the case of a non-intersecting
loop, $\alpha$, the bra is, for every $\omega$,  an eigenstate
of the operator corresponding to
$Q[\omega , \tilde{E}^{ai} ] $.  The action of the operator is given by,
\f
<\alpha | \hat{Q}[\omega ] = {l_{Planck}^2 \over 2}\, \
   \oint_\alpha ds\, \ |\dot{\alpha}^a \omega_a(\alpha(s))|\,\
   < \alpha| .
\ff

A second nonlocal operator that can be defined as a quantum
operator is the area of any surface.   Given a surface $\cal S$, there is
a function on the kinematical phase space that is its area, it is given
by,
\f
{\cal A}[{\cal S},q] = \int_{\cal S} \sqrt{\tilde{\tilde{q}}^{ab}n_an_b}
\ff
where $n^a$ is the unit normal of the surface.  The problem is how to
turn
this into an operator when there is no operator for the metric
$\tilde{\tilde{q}}^{ab}$?  There is a solution, which is the following.
Let me represent the surface by a distributional one form,
$\pi_a^{\cal S}$ which
is given by,
\f
\pi_a^{\cal S} (x) = \int d^2S^{bc}(\sigma ) \
delta^3 (x ,{\cal S} (\sigma ))
\epsilon_{abc},
\ff
where $\sigma$ are  coordinates on the surface and $\epsilon_{abc}$ is
the inverse of the Levi-Civita density.   I then can consider
the expression
\f
A({\cal S}, \tilde{E}) = Q(\pi, \tilde{E})
\ff
It is not difficult to show that this is equal to the
area of the surface
given by the metric by (27).  To show this, we demonstrate that an
equivalent expression, when the frame fields are smooth, is given by
\f
A(\pi ,\tilde{E} ) = \lim_{N \rightarrow \infty} \sum_{N=1}^N
\sqrt{A^2_{approx} [{\cal R}_i]}
\ff
where space has been partitioned into $N$ regions
${\cal R}_i$ such that
in the limit $N \rightarrow \infty$ the regions all shrink to
points.  Here,
the
observable that is measured on each region is defined by,
\f
A^2_{approx} [{\cal R}] \equiv \int_{\cal R} d^3 x \int_{\cal R} d^3y
T^{ab} (x,y) \pi_a (x) \pi_b (y)
\ff
Here $T^{ab} (x,y)$ is a loop operator that is quadratic in the frame
fields $\tilde{E}^{ai}$.  It is constructed in the following way.  Pick a
background Euclidean metric and use it to define, for
every two points, $x$ and $y$, in the spatial manifold
$\Sigma$, a circle, $\gamma_{xy}$, such that
$\gamma_{xy}(0)=x$ and $\gamma_{xy}(\pi )=y$ and such
that in the limit that $y$ approaches
$x$, the circles shrink to the point $x$.  Then, define
\f
T^{ab}(x,y)= {1\over 2} Tr\left [({\cal P} \, \exp
        {G \int_{y}^x A_a d\gamma^a_{xy}})\, \tilde{E}^a(x)\, ({\cal P}
        \exp {G \int_x^{y} A_a d\gamma^a_{xy}})\,
        \tilde{E}^{b}(y) \right ].    \nonumber
\ff
To show the equivalence between (29) and (30), we start with (29) and
regulate it by means of a point splitting procedure  by
introducing, with respect to the
background euclidean coordinate system,  a set of test fields
$f_{\epsilon} (x,y)$ by
\f
f_{\epsilon}(x,y) = {1 \over \epsilon^3 }
\theta [{\epsilon \over 2}-|x^1-y^1 |]
\theta [{\epsilon \over 2} -|x^2-y^2 |]
\theta [{\epsilon \over 2}-|x^3-y^3 |].
\ff
In these coordinates
\f
\lim_{\epsilon \rightarrow 0} f_{\epsilon} (x,y) = \delta^3 (x,y)
\ff
We can then write
\f
A(\pi , \tilde{E}) = Q(\pi,\tilde{E}) =
\lim_{\epsilon \rightarrow 0} \int
d^3x
\sqrt{\int d^3 y \int d^3z T^{ab}(y,z) \pi_a (y) \pi_b (z)
f_\epsilon(y,x) f_\epsilon (z,x)}
\ff
When the expression inside the square root is slowly varying in
$x$ we can
re-express it in the following way.  We divide space into regions
${\cal R}_i$ which are cubes of volume
$\epsilon^3$ centered on the points
$x_i = (n\epsilon , m\epsilon , p \epsilon )$ for $n,m,p$ integers.
We then
write,
\begin{eqnarray}
A(\pi , \tilde{E}) &= &\lim_{\epsilon \rightarrow 0} \sum_i \epsilon^3
\left [ \int d^3y \int d^3z T^{ab}(y,z) \pi_a (y) \pi_b (z)
f_\epsilon(y,x_i) f_\epsilon (z,x_i)\right ]^{1 \over 2}  \nonumber  \\
&=&\lim_{N \rightarrow \infty} \sum_{N=1}^N \sqrt{A^2_{approx} [{\cal
R}_i]}
\end{eqnarray}

If we now plug into these expressions the distributional form (28) it is
straightforward to show that
\f
A(\pi_{\cal S} , \tilde{E})= \int_{\cal S} \sqrt{h}
\ff
where $h$ is the determinant of the metric of the two surface, which is
given
by $h=\tilde{\tilde{q}}^{ab}n_a n_b$ where $n^a$ is the unit normal
of the
surface.

It is not difficult to show that that starting from the expressions (30)
and (31) we
may construct a quantum operator for the area of a surface
$\cal S$ in the loop representation.  We can show that the expression
(30) is equivalent to an expression in which the surface is
partitioned into
$N$ subsurfaces ${\cal S}_I$, $I=1,...,N$.  We then write
\f
 {A}_S = \lim_{N \rightarrow\infty}\, \, A_{\rm appr}^2 [{\cal S}_I].
\ff
where $  A_{\rm appr}^2 [{\cal S}_I]$ denotes an approximate expression
for the area of the subsurface, which is defined by
\f
 A_{\rm appr}^2 [{\cal S}_I] \equiv \int_{S_I} d^2S^{bc}(x)
 \epsilon_{abc}
   \int_{{\cal S}_I} d^2S^{\prime b'c'} (x') \epsilon_{a'b'c'}\,
 T^{aa'}(x,x')
\ff

This last expression may be written as a quantum operator, by writing
an operator for $T^{aa'}(x,x')$ .  This can
be done, but, as the action is a bit complicated, I do not give it here.
It may be found in \cite{carlolee,review-abhay,review-carlo,review-lee}.
The result is that the limit (38)
may be taken on any loop state, leading to a  final expression that
is finite
and independent
of the background structure that went into the
definition of the loop operator.  The result, for non-intersecting
loops $\alpha$ is\cite{review-lee,weave},
\f
<\alpha | \hat{A}[{\cal S}] = { l_{Planck}^2 \over 2}
I^+[{\cal S}, \alpha ]
<\alpha | .
\ff
Here $I^+[{\cal S}, \alpha ]$ is the positive, unoriented, intersection
number, which counts (independent of orientation) the number of
intersections of the loop with the surface.

If the loop $\alpha$ has intersections the action of the operator is
more complicated, but it is still finite and background independent.
Details are given in \cite{review-lee}.

The third observable that can be constructed in this way
is the volume of
any
region.  It is described in \cite{review-lee}

\subsection{Quantization of areas and volumes in the discrete
representation}

The result (40) says that the bras in the loop representation
are, at least for
intersecting loops, eigenstates of the operator that measures
areas.  This
does not mean that, in general, there are normalizable states that are
eigenstates, this can only be the case if the inner product is defined in
such a way that there is a state which is the hermitian conjugate of the
bra $<\alpha|$ which is a normalizable state and if the area operator is
hermitian in that inner product.

In general these conditions will not be satisfied, for example,
there are no
normalizable eigenstates of the area operator in the Fock representation
of linearized quantum gravity.  But in the context of discrete
representations an inner product can be defined by the imposing the
condition that the inner product be chosen such that the
area operator is hermitian so that its eigenstates comprise  an
orthonormal basis.  For nonintersecting loops this is given
by (22), for intersecting loops it is more
complicated \cite{review-lee}.

With such an inner product, we may say that area is quantized
in the discrete representation, because the
spectrum of the operator that measures area is discrete.  This
spectrum consists, first of all, of the
eigenvalues $N l_{Planck}^2 /2 $, for every nonnegative integer
$N$.  There is also another discrete sequence of  eigenvalues
corresponding to eigenstates that are labeled by intersecting
loops, these
are described in \cite{review-lee}.

In the same representation, the volume operator turns out also
to have a
discrete spectrum.  The basic action of the volume operator in this
representation turns out to be to annihilate the states $|\alpha>$
associated with nonintersecting loops $\alpha $ and to rearrange the
routings through the intersections of the intersecting loops.
That is, with
each intersecting loop, $\alpha $, one can associate a finite
dimensional
subspace of the state space which is spanned by the loops
which have the
same support as $\alpha$ but differ as to how the loops are
routed through
the intersection points.  The action of the volume operator
is then to
induce a finite dimensional matrix in each such subspace that
rearranges
the routings then multiplies by
$l_{Planck}^3$. Its non-zero eigenvalues are given by
$l_{Planck}^3$ times
the eigenvalues of these  finite dimensional matrices.

\subsection{The correspondence principle:  Existence of states which
approximate classical metrics at large scales}

We are used to describing the classical limit of quantum theories in
situations in which there is a background metric against
which to measure
distance intervals.  It is not a completely trivial problem to understand
what it means to take the classical limit in a non-perturbative quantum
theory of gravity, in which there is no background metric.
To do this we need to first understand two simple points.
First,   in pure quantum general relativity the classical limit
is a limit of
large distances.  This is because the theory has only one dimensional
parameter, the Planck length,
$l_{Planck}=\sqrt{\hbar G/c^3}$.  This obviously goes to zero as
$\hbar \rightarrow 0$.  It is perhaps also significant that it
is the Planck
area that is proportional to $\hbar$, this perhaps is the
reason why length
intervals are not defined in the quantum theory, while
areas are both
defined and are quantized.

The second thing to be understood is that without a
classical metric we
don't know what distance and area means and so we cannot tell which
intervals are small or large compared to the Planck scale.

 Because of these two points, it is easiest to express the
 classical limit
in a way that may seem backwards, as
follows\cite{weave,review-carlo,review-lee}.
Given any classical metric
$h_{ab}$, whose curvatures are small compared to the Planck scale, we
seek a quantum state $|\Psi >$ which has the property that
it is an eigenstate of the operators $\hat {Q}[\omega ]$ and
$A[{\cal S}]$, and where, for every
one form $\omega$ which is slowly varying with respect to the metric
$h_{ab}$ and every surface which has small extrinsic curvatures, again
with respect to $h_{ab}$ (where, again these area measured with respect
to the Planck scale) we have
\f
\hat Q [\omega ] |\Psi > = \left (  Q(\omega , h ) +
O( l_{Planck} |\bigtriangledown
\omega | ) \right ) |\Psi > ,
\ff
and
\f
\hat A [{\cal S}] |\Psi > = \left ( A[{\cal S}, h] +
O \left ( {l_{Planck}^2 \over A[{\cal S}, h]}
\right ) \right ) |\Psi > .
\ff

Thus, the eigenvalues are required to give
back the corresponding values for the metric $h_{ab}$ up to terms
that are small measured in Planck units.
When these conditions are
satisfied we say that $|\Psi >$ is a semiclassical state that
approximates
the metric $h_{ab}$.

In the loop representation we call states that have this property
weaves, because it can be satisfied by loop states
$|\Delta >$, where
the multiloop $\Delta$ consists of many small loops which are
arranged so that, through every surface, ${\cal S}$, as
described above,
approximately one line of a loop pierces $\cal S$ per half
Planck area of the surface, measured in the metric $h_{ab}$.
It is easy to give examples of such states; a particularly
simple one, for the case that the metric
$h_{ab}$ is flat,  is constructed as follows \cite{weave}.

We use  $h_{ab}$, to  introduce a
random distribution of points on $\Sigma = R^3$ with density $n$.  This
means
that in any given volume $V$ there are $nV(1+ {\cal O}(1/\sqrt{nV}))$
points. We center a circle of radius $a = (1/n)^{1\over 3}$ at each of
these points, with a random orientation.   Again, the notion of
a random orientation is defined with respect to $h_{ab}$.  We call this
whole  collection of circles $\Delta$.

It is now straightforward to show that  $|\Delta >$ is an eigenstate
of $\hat Q [\omega ]$ and $A[{\cal S}]$.  However the conditions
(41) and (42) are only satisfied if the density $n$ is
chosen so that\cite{weave},
\f
a =
\sqrt{\pi/2}
\, l_p.
\ff

\subsection{The necessity of discrete structure at the Planck scale}

This last result (43) means that if we require that
the state $|\Delta >$
approximate the classical metric $h_{ab}$, when we measure it with
operators that average the metric information over scales that are
large in Planck units, it is necessary
that the state have discrete structure at the Planck scale, where,
in both cases, what we mean by the Planck scale is determined by
$h_{ab}$.  This is a direct consequence of the fact that we were able
to construct non-local operators to measure the metric information
that are finite and background independent.  This result can be
generalized by considering families of loops that generalize
$\Delta$ by being described by more parameters.  In each case it
is found that there is one combination of parameters that is fixed
to be a certain exact multiple of the Planck scale.  The
other combinations
of parameters with dimensions length are also restricted to be on the
order of the Planck scale, so that the
requirements on the orders of the errors in (41) and (42) are
satisfied\cite{weave}.

This means that in nonperturbative quantum gravity, at least in the
formulation I am describing here, it is possible to have states that
are semiclassical on large scales.  However, there are no states that
are semiclassical on the Planck scale.

This completes my discussion of the kinematical level of the theory.
I now turn to results concerning diffeomorphism invariant states.

\subsection{Complete solution of the diffeomorphism constraints}

We have defined the action of diffeomorphisms on states in the loop
representation by (23) and (24).  Using these, we define the
space of diffeomorphism
invariant states, ${\cal V}_{diffeo}$, to be those loop states
that satisfy,
\f
\Psi [\alpha ] = \hat U (\phi ) \Psi [\alpha ] =
\Psi [\phi^{-1} \circ \alpha ]
\ff
for all elements of the connected
component\footnote{There are two ways to treat the
large diffeomorphisms: as symmetries or as part of the
gauge group.  I do not discuss this issue here.} of
the diffeomorphism group of
$\Sigma$.

The definition (44) of diffeomorphism invariant states may
be compared with a
similar condition in the metric representation, which says
that the quantum states are functions of the three geometry,
which are defined to be diffeomorphism equivalence classes
of three metrics.  The difference is that the diffeomorphism equivalence
classes of loops are countable\footnote{Assuming certain
mild conditions on the finiteness of the components and the
intersections.}\cite{knots} and a great deal is known about their
classification.  If we denote by $\{ \alpha \}$ the diffeomorphism
equivalence class of the loop $\alpha$  (also known as its
knot or link class) the
condition (44) means that\cite{carlolee}
\f
\Psi [\alpha ] = \Psi [\{ \alpha \} ] .
\ff
Because the knot classes are countable the space of diffeomorphism
invariant
states, ${\cal V}_{diffeo}$,  has a countable basis, which are the
characteristic states
of the knot classes.  These are,
\f
\Psi_{\{ \gamma \} } [ \{ \alpha \} ] =
\delta_{\{ \alpha \} \{ \gamma \} } .
\ff

We can make ${\cal V}_{diffeo}$ a Hilbert space by imposing
an inner product.
The simplest possibility is one in which these characteristic
states are orthogonal,
that is if we chose a basis of bra states $<\{ \alpha |$ such that
$\Psi[ \{ \alpha \} ] = <\{ \alpha \} |\Psi >$ and we chose
the inner product so that
$<\{ \alpha \} |^\dagger = |\Psi_{\{ \alpha \} } >$,
we have
\f
<\{ \alpha \} |\Psi_{\{ \gamma \} } > =
\delta_{\{ \alpha \} \{ \gamma \} }
\ff
This is almost certainly not the right inner product for
general relativity, because it
corresponds to a reality condition in which the connection
is real.  However, it may be
useful as a technical device to bound limits in certain calculations.

It should be mentioned there is a diffeomorphism invariant
theory for which (47)
is  the physical inner product.  This is the
Husain-Kuchar model\cite{husainkuchar},
which is a limit of general relativity in which the speed of
light has been taken to
infinity\cite{carloobserves}.  In the classical version of this
theory, all
physical evolution has been frozen, and all solutions are static.
Because of this the theory has no Hamiltonian constraint-it has
only the gauge
and diffeomorphism constraints.

The Husain-Kuchar model is a very interesting model because it
is a three plus one
dimensional diffeomorphism invariant theory that has an infinite
number of physical
degrees of freedom.  It is solved to the extent that the exact
state space and inner
product have been constructed.  The theory needs to be completed by
the construction
of a sufficient number of diffeomorphism invariant observables,
represented by operators on ${\cal V}_{diffeo}$, on which the
interpretation can be grounded.   As there
are only the spatial diffeomorphism invariant constraints,  this
problem is significantly
easier than in the case of the full theory.  The Husain-Kuchar model
is a very useful laboratory to study those  problems of the
interpretation of
diffeomorphism invariant quantum field theory that are
{\it not} related to the problems of time and time reparametrization
invariant
observables.

\subsection{Some finite diffeomorphism invariant operators}

In fact, a small number of diffeomorphism invariant observables
can be directly written down.  I will describe here one of them,
which is closely related to the area observable I described in
subsection 3.8 above.  The idea is to introduce a dynamical
field whose configuration can define a surface.  The area of that
surface will
then be a diffeomorphism invariant quantity\footnote{The idea of
using matter
fields to define physical and diffeomorphism invariant observables
is an old idea,
which goes
back at least to a paper of DeWitt \cite{Bryce-matter}.  It has been
recently revived
 \cite{carlo-matter,karel-matter}.}.

One way to do this is to couple an antisymmetric
tensor gauge field to gravity\cite{lee-matter}.  This is a two
form, $C_{ab}=-C_{ba}$ subject to a gauge transformation generated by
a one form $\Lambda_a$ by,
\f
\delta C_{ab} = d\Lambda_{ab} .
\ff
It's field strength is a three form which is denoted $W_{abc}=dC_{abc}$.
In the Hamiltonian theory its conjugate momenta is
given by $\pi^{ab}=-\pi^{ba}$ so that
\f
\{ C_{ab} (x) , \pi^{cd} (y) \} = \delta^{[c}_a \delta^{d]}_b
\delta^3 (y,x) .
\ff
The gauge transform (48) is then generated by the constraint
\f
G=\partial_c \pi^{cd} =0
\ff

Any two dimensional surface $\cal S$ defines a  distributional
configuration of
the $\pi^{ab}$ by equation (28), where we now want to understand
the field $\pi_a$ in that equation as being the dynamical field
dual to $\pi^{bc}$ by $\pi_a =
\epsilon_{abc}\pi^{bc}$.
 These distributional configurations are solutions to the gauge
 constraint (50) and
 I thus know of no reason it cannot be considered to be an allowed
 configuration of the classical field.

We can then interpret equation (29) as the definition of a
diffeomorphism invariant observable by reading it as  a function
of a dynamical  $\pi_a$ field and the gravitational field.
It has the interpretation that when the $\pi^{ab}$ field
defines a surface through  a distributional configuration by
equation (28), it gives the area of that surface.

This observable can be promoted to an operator if we also quantize
the $C_{ab}$ field.  This can be done by constructing a surface
representation to represent it, completely analogous to the loop
representation.  To do this we introduce a surface
observable,
\f
T[{\cal S} ]= e^{k  \int_{\cal S} C}
\ff
associated to every closed surface $\cal S$.  The $k  $ is a
free constant with dimensions of inverse action\cite{lee-matter}.
The algebra we will
quantize is then the surface algebra
\f
\{ T[{\cal S}] , \pi^{bc} (x) \} =k \int d^2{\cal S}^{bc} (\sigma )
\delta^3 (x, {\cal S}(\sigma ) )T[{\cal S}]
\ff
We can then construct a representation of this algebra in which
states are functions of surfaces $\Psi [{\cal S}]$.
We then define the representation by\cite{lee-matter},
\f
\hat{T}[{\cal S}^\prime ] \Psi [ {\cal S}] =
\Psi [ {\cal S}^\prime \cup {\cal S}]
\ff
and
\f
\hat{\pi}^{ab} (x) \Psi [{\cal S}] =
\hbar k \int d^2 {\cal S}^{ab} (\sigma )
\delta^3 (x, {\cal S}(\sigma ) ) \Psi [{\cal S}]  .
\ff
To represent the coupled $C_{ab}$-gravity system we take the
direct product of this state space with the loop representation
for quantum gravity.
The states are then functions, $\Psi [\alpha , {\cal S}]$,
of loops and surfaces.  We may introduce a set of bra's,
$<\alpha , {\cal S}|$, labeled by loops and surfaces so that
$\Psi [\alpha , {\cal S}] =<\alpha , {\cal S}|\Psi > $.

We may then impose the diffeomorphism constraints, suitably
extended to the coupled
Einstein-$C_{ab}$ system\cite{lee-matter}.  I
will not give the details here, the result is that the diffeomorphism
invariant states may be constructed and they are functions of the
diffeomorphism equivalence classes of loops and surfaces.  Denoting
these classes by $\{ \alpha , {\cal S} \}$, the
diffeomorphism invariant
state space then consists of functions of the form
\f
\Psi [\{ \alpha , {\cal S} \} ] =<\{ \alpha , {\cal S}\} |\Psi >  .
\ff
We then want to express the area observable (37) as a
diffeomorphism invariant operator and show that it does indeed
measure areas.  It is straightforward to show that the bras at
the kinematical level,
$<\alpha , {\cal S}|$, are, for nonintersecting loops $\alpha$,
eigenstates of
the operator $\hat{A}$.
This operator may be constructed by using the expressions (30) and
(31) as the definition of a
regularization procedure, in the usual way\cite{review-lee}.  As the
regularization breaks diffeomorphism invariance, this calculation must
be done at the kinematical level.  A straightforward calculation
shows that
\f
<\alpha , {\cal S}| \hat{A}^2_{approx} [{\cal R}] =
({ \hbar kl_{Planck}^2 \over 2 })^2 I[\alpha , {\cal S} \cap {\cal R} ]^2
<\alpha , {\cal S}|
\ff
where ${\cal S} \cap {\cal R} $ means the part of the surface that lies
inside the region.  It then follows from (30) that
\f
<\alpha , {\cal S}|  \hat{A} = { \hbar k l_{Planck}^2 \over 2 }
I^+[\alpha , {\cal S}]  <\alpha , {\cal S}| \nonumber
\ff
This may be compared
with (40), we see that the only difference is that now that the
surface is dynamical it is specified by the state and not by the operator.   In
addition, we see that if we want agreement between the units
measured by this and the kinematical area operator we must pick
$k=1/\hbar$.

The action of $\hat A$ can be lifted to the space of
diffeomorphism invariant
states, giving us,
\f
<\{ \alpha , {\cal S} \} |  \hat{A} = { \hbar k l_{Planck}^2 \over 2 }
I^+[\{ \alpha , {\cal S} \} ]  <\{ \alpha , {\cal S}\} | \nonumber
\ff
We have thus defined a diffeomorphism invariant operator that
assigns to the surface an area which is given by
$\hbar k l_{Planck}^2 / 2 $ times the number of intersections
of the loop with the surface.
Thus, we see that the same techniques that gave us finite and
background independent kinematical operators work to give us
finite operators acting on diffeomorphism invariant states.

If we use the inner product (47) of the Husain-Kuchar model on
the space of diffeomorphism invariant states, suitably extended
to include the coupling to the $C_{ab}$ field \cite{lee-matter},
we see that $\hat A$ is a hermitian operator and that its
spectrum is quantized.  Thus, we see that the technology we have
been developing allows us to derive a prediction from a $3+1$
dimensional diffeomorphism invariant quantum field theory.   This
is that the area of any two dimensional surface is quantized in units
of the Planck area.  While that theory is a model, which corresponds
classically only to a limit of full general
relativity, this is an encouraging result.  Moreover, it does not seem
impossible that with the addition of structure corresponding to clocks
it will be possible to extend the $\hat A$  observable to the full
physical case, and that we will find that
this prediction stands in full quantum general relativity.

It is known that a few additional diffeomorphism invariant operators
can be constructed in this way, in either the case of pure
gravity or gravity coupled to matter.  Examples of
these\cite{review-lee} are the volume of the universe and the areas of
maximal surfaces
(when $\pi^2$ of the spatial manifold is non-trivial.)  Of course,
there
must be an infinite number of diffeomorphism invariant observables,
it is still
an open problem to show that these techniques allow the construction
of an infinite
number of such
operators.

\subsection{Connection between finiteness and diffeomorphism
invariance of operators}

All of the diffeomorphism invariant operators which have so far been
constructed are also finite.  We may ask whether finiteness is a
general
property of diffeomorphism invariant operators constructed
nonperturbatively.
There is a general argument that this is the case; which I
would like
to sketch here.

The argument begins with the assumption that any diffeomorphism
invariant
operator that will exist in a nonperturbative quantum theory must be
constructed through a regularization procedure.  All such
procedures which
are known require that one introduce both a background metric
and a regulator
scale.   This is
necessary, because the scale that the regularization parameter
refers to must be described in terms of some metric and, since
none
other is available, it must be described in terms of a background
metric or coordinate chart
introduced in the construction of the regulated operator.  Because
of this, the dependence of the regulated operator on the
cutoff parameter is related to its dependence on the
background metric.  This can be formalized into a kind of
renormalization group equation \cite{carloun}.  When one takes the
limit of the regulator parameter going to zero one isolates the
nonvanishing terms.  If these have any dependence on the
regulator parameter (which would be the case if the term
is blowing up) then it must also have a dependence on the
background metric.  Conversely, if the terms that are nonvanishing
in the limit the regulator is removed have no dependence on the
background metric, {\it they must be finite.}

This point  has   profound
implications
for the whole discussion of finiteness and renormalizability of
quantum gravity theories.  It means that any nonperturbative and
diffeomorphism invariant construction of the observables of the
theory must be finite.  A particular approach could fail in that
there
could be no way to construct the diffeomorphism invariant
observables as quantum operators.  But if it can be done, without
breaking diffeomorphism invariance, those operators will be finite.

\subsection{Exact physical states of the quantum gravitational field}

We come finally to  the physical state space,
${\cal V}_{phys}$, which consists
of those  states which are solutions to all of the
constraints of quantum general
relativity, including the Hamiltonian constraint.
Although it is logical to put the
discussion of these last, this is not the order in which the subject
actually developed.
The discovery that the Hamiltonian constraint could be exactly
solved was made first\cite{tedlee}; later the loop representation
was invented to solve the diffeomorphism constraint \cite{carlolee}.
At the time the first solutions to the Hamiltonain
constraint were found, it was possible to imagine that
the existence of
an infinite dimensional space of exact solutions to the
Hamiltonian constraint might be some kind of spurious result,
having nothing to do with physics.  Now, six years later,
after the development of the loop representation, and after
we have understood
how the discreteness of the quantum representation acts, at
the kinematical and diffeomorphism invariant level, to allow
the existence of finite, background independent operators and
discrete structure at the Planck scale, and after we have further
understood the role of this discreteness in assuring the existence
of the classical limit,
that the Hamiltonian constraint can be solved in this way seems
much more natural.

In order to define its action in any representation,
the Hamiltonian constraint
must be regulated.  In the literature there are four
different proposals for how
to carry out this regularization, due to Rovelli and
the author\cite{carlolee},
Gambini \cite{gambini-newloop}, Blencowe
\cite{miles} and Bruegmann and Pullin \cite{BP-onC}.
These are now understood to be equivalent, at least
when acting on a certain class of states \cite{BP-onC}.

The result of one of these regularization procedures is a sequence
of well defined operators ${\cal C}^\delta $ in ${\cal V}_{kin}$,
for $\delta > 0$.  A solution to the quantum Hamiltonian constraint
is then taken to be one such that,
\f
\lim_{\delta \rightarrow 0} \left ( {\cal C}^\delta |\Psi > \right ) = 0
\ff
The physical states are then taken to be those states that are
simultaneous solutions to this condition and the
diffeomorphism constraint (44).

I would like to make several comments about this condition.

a)  There is no necessity that the limit $\lim_{\delta \rightarrow 0}
{\cal C}^\delta$ define an operator in ${\cal V}_{kin}$.  One could
construct such an operator by a renormalization procedure
in which the operator
was multiplied by the appropriate
power of $\delta$ as the limit is taken.  But, it is not clear
what use this
would be.  The operator, in any case, must vanish on the space
${\cal V}_{phys}$
which we are interested in, and on ${\cal V}_{kin}$, for the
reasons we discussed in section 3.7, it will not be diffeomorphism
invariant.

b)  The Hamiltonian constraint, even before regularization,
is not diffeomorphism
invariant, as it is the integral of an arbitrary density with
the local function of the fields.  Thus, it does not define an
operator in ${\cal V}_{diffeo}$.  It would be very
interesting to have an expression for the projection of the
diffeomorphism constraint
into that space.  One could do this, for example, by finding an
infinite set of
functions on the classical phase space that vanish on the same
surface as ${\cal C}$,
but which are diffeomorphism invariant, and then represent those
as quantum operators.

c)  An issue that is often raised is the question of whether
the algebra of the
constraints, quantum mechanically, has an anomaly which
would prevent the existence of
simultaneous solutions to all of them.  In fact, as I am about
to describe, we know of
infinitely many simultaneous solutions, so there can be no
anomalous terms in the algebra which is
proportional to the identity operator in the state space.
Furthermore, as we have just
remarked, the Hamiltonian constraint does not define a good
operator on
${\cal V}_{kin}$ unless we further break diffeomorphism
invariance by the addition
of a renormalization procedure, so it is not clear exactly
what condition to ask from
our quantum operator algebra.  However, it would still be
interesting to know what
the algebra of the regulated operators is like; this problem
is presently
under study\cite{gambini-pullin-personal}.

d)  In the condition (59), the limit is taken in the pointwise
topology, which
means that the limit must vanish when taken over every point of
the loop space.
As the physically meaningful inner product is constructed on the
space of solutions
to the
constraint, this is sufficient as long as that solution space is
large enough.
However, it would be surprising if the limit could also not be
expressed in terms of a
Hilbert structure on ${\cal V}_{kin}$.  I believe that this can
be done, but the details have not been
worked out.

The result of the regularization procedures is that the
Hamiltonian constraint can be
given a kind of geometrical interpretation when acting in
the loop representation.
First of all, acting on states that have support only on
loops without
intersections, the limit (59)
vanishes\cite{tedlee,carlolee,gambini-newloop,BP-onC}.
Thus, the action of the operator is, in the limit, only
sensitive to the behavior
of the state at intersecting loops.  At an intersection,
the action of the operator
consists of
two
parts:
first,
a rearrangement
of the routings through the intersection and second, a
loop derivative taken at the intersecting
points\cite{carlolee,gambini-newloop,BP-onC}.

At the present time, there are several different sectors of
solutions to the physical state space which have been
explicitly constructed.
First, as I have just mentioned, any state that has support
on only nonintersecting
loops is a solution.  This is
an infinite dimensional space;  among these is a state
corresponding to every
invariant of nonintersecting links\cite{carlolee}.
Then, there are two different sectors of
states which have been constructed which have support on
intersecting loops.  The
first consists
of characteristic states, which have support on only a finite
number of diffeomorphism
equivalence classes of intersecting loops.  These have
been constructed for intersections
at which two \cite{tedlee}, three \cite{viqar-intersects},
four and five \cite{berndtjorge-intersects} lines meet.  Then,
very recently, a new sector of physical states has been
discovered by Bruegmann, Gambini
and Pullin, which are closely related to the Jones
polynomial\cite{BGP}.

At present, the study of exact physical states is ongoing,
and there are a number
of open problems.  It is clear that the full set of
solutions is not known, and
nothing is known about the relationship between
the different sectors of the
solution space.
 Most importantly, it has not been established the
 extent to which the
 known types of solutions characterize the general
 solution to the constraints.

Of course, the construction of the theory is not
complete without further
elements, particularly the physical observables and
the physical inner product.
I will discuss the open problems in the next section.
For the remainder of
this section, I
would like to discuss a number of results concerning the
application of these
nonperturbative methods to models which are simpler than full $3+1$
dimensional quantum gravity.

\subsection{Application of the loop representation to 2+1 gravity}

As Witten first pointed out, quantum general relativity in
$2+1$ dimensions
is exactly solvable because for each spatial topology one has,
after the
solutions of the constraints, a finite dimensional phase
space \cite{2+1-witten}.
Thus, although the model has only a
finite number of degrees of freedom, it provides a good test of many of
the ideas and methods that were originally developed in
the $3+1$ dimensional case.
The theory can be completely solved in using both the connection
representation\cite{2+1-witten,review-abhay}
and the
loop representation\cite{2+1-us,review-abhay,2+1-me}.
The physical operators can be constructed, and they turn
out to be closely related to the loop operators (2) and (4).
The physical inner product is also easily constructed.

In addition, as Carlip has shown in a very elegant series of
papers\cite{Carlip-time},
the difficult problem of time can be resolved completely in
this model, along the lines
proposed by Rovelli \cite{Carlo-time}.

A case which lies intermediate in difficulty between this case
and the full
$3+1$ case is that of $2+1$ gravity coupled to matter.  In
particular, with
the addition of one scalar field one has a model that can
also be interpreted
as $3+1$ gravity with
one killing field\cite{2+1-killing}.

Recently Ashtekar and Varadarajan have studied this model, and have
found a number of
interesting results at the classical level\cite{abhaymadhavan}.
The most important of these is that it is possible to define a
notion of asymptotic flatness, such that the energy is
bounded both from above and from below.  Using very different
methods, Bonacina,
Gamba and Martellini
have shown that this theory is also perturbatively
renormalizable\cite{maurizio-2+1}.
In addition, there have been two interesting papers in
which such systems are treated
in the loop
representation;  which treat the coupling of Maxwell fields to $2+1$
gravity\cite{2+1-maxwell}, and $3+1$ gravity with one
killing field\cite{BP-1kf}.

\subsection{Application of the loop representation and new variables
to two killing field reductions}

Another very interesting model is general relativity with two
killing fields,
as this is known to be an integrable system with an infinite
number of degrees of
freedom.  This system has been studied using the new variables
and the loop
representation,
and a number of interesting results were obtained\cite{viqarlee}.
The key open problem in this area is to represent the generators
of the Geroch group as canonical transformations, generated by an
infinite dimensional algebra physical observables of the
model.  Some
very interesting partial results in this direction have been obtained by
Torre\cite{charles-2kf}.

\subsection{Nonperturbative quantization of  the Bianchi models}

A last type of model I would like to mention is the class of
Bianchi cosmologies.
These are finite dimensional model cosmologies, which offer
good laboratories
for ideas about quantum gravity and quantum cosmology.
Most of these have not
been solved,
in spite of the fact that they have only a few degrees of freedom;
these models
thus serve as a reminder that in quantum gravity, as in ordinary
quantum mechanics,
finite dimensional does not imply solvable.

It would be very interesting to be able to solve these models, through
approximation
methods if not exactly.  There are a number of interesting
results concerning
them  which employ Ashtekar's
variables\cite{kodama1,kodama2,abhayjorge}.
Among these is the construction of a
set of exact
states for the Bianchi IX model by Kodama \cite{kodama2}
but, as in the full theory,
little is known about the physical observables or the
inner product in this model.

\section{What are the key open questions?}

In quantum gravity, it is safe to assume that any
important problem is
difficult, until the occasion of some progress provides
evidence to the contrary.
Indeed, in the development of the work I have been describing
here, almost every
result came out in
a surprising way.  Actually, once understood correctly,
most of the results are not very
difficult; the key seems to be to
ask the question in precisely the right way.

Given this, the key open questions are simply how to construct
those elements
of a quantum field theory that are, so far, missing.  I give
here a list of them;
more detailed discussions about each of them may be found in the
reviews \cite{review-abhay,review-carlo,review-lee,time-evolution}.

\subsection{How can we construct the physical observables?}

As I mentioned above, the problem of physical observables in
quantum gravity is difficult partly because there is already
a problem in the classical theory.  The problem can be stated
this way:  any classical observable in general relativity,
with cosmological boundary conditions, must be a constant of
motion.  This is because to be invariant under diffeomorphisms
it must commute with the Hamiltonian constraint, but in the
cosmological case the Hamiltonian is proportional to a linear
combination of constraints.  This must be; were there a
meaningful nonvanishing Hamiltonian it would be meaningful
to ask how fast the universe is evolving, so that evolutions
that differed only by the rate at which time progressed would
be physically
distinct.  As there can be no clock outside the universe,
this cannot be meaningful.

Thus, the problem of physical observables is closely
connected with the notion of time.  As such, it is one of
those great problems that are both conceptually and
mathematically profound .

At the present time, a rather large number of ideas are
being studied with an aim towards solving
this problem.  I list here the ones I am aware of, with references.

i)  Coupling the theory to  matter, and using this matter
to provide a system
of clocks with which to make observables meaningful.  This
is a very old
idea\cite{Bryce-matter},
recently it has received a lot of
attention\cite{carlo-matter,carloobserves,lee-matter}.

ii)  Imposing asymptotically flat boundary conditions,
which provide an
observer and a classical clock at infinity.  The problem
with such an
approach is that spatial infinity is, in a certain sense,  too
far away, and only a limit number of observables,
corresponding essentially to the globally conserved quantities,
may be
defined there.  Still, this is undoubtably worth doing, and some recent
results of Baez are very interesting in this regard\cite{Baez}.
The key open
problem with this approach is
to show that, with respect to the correct inner product, the
quantum Hamiltonian is
bounded from below.

iii)  Imposing some other kind of boundary conditions, which may
allow more observables
to be introduced.   One such idea, due to Crane, is to define
observables on two
dimensional surfaces, and use conformal field theory thereby
as a kind of measurement
theory for quantum cosmology\cite{louis-idea}.  Another related
idea involves choosing boundary conditions so that a Chern-Simon
theory is induced on the boundary \cite{review-lee}.

iv)  Studying certain limits of the theory,
where observables can be constructed
\cite{carloobserves,Gtozero}.

v)  Finding an approximation scheme, such as a strong coupling
expansion or a new
form of perturbation theory, that will allow observables
to be constructed systematically.

vi)  Modifying the interpretative rules of quantum cosmology,
so as to make
the problem easier to solve\cite{Julian-time}.

vii)  Construct observables that are associated with global
properties of the
configuration of the gravitational field.  This has led,
during the last year,
to the construction of the only explicit examples yet discovered of
observables of the pure gravitational field\cite{newobservables} .

ix)  Construct a superposition of exact states that corresponds,
in the semiclassical sense described above, to Minkowski
spacetime.  Small perturbations on this state should then
correspond to gravitons traveling on Minkowski spacetime, at
least for
long wavelengths.  To show this one can construct a map
from a sector of the Fock space of linearized quantum gravity
into a subspace the space of exact physical states.  This map
then can be used to construct an approximate interpretation
of the exact
states in that subspace.  There is some preliminary evidence
that this map exists\cite{zegwaard,junichicarlo}.

\subsection{How can we construct the physical inner product?}

The last structure that is necessary to do physics is the
inner product.
This problem is closely connected to the problem of the
observables because,
in the absence of a global Poincare covariance, the inner
product must be picked
by the requirement
that a complete set of real classical observables are represented
by self-adjoint
operators.   Further, since the observables are constants of
motion, the problem of
determining the inner product is a dynamical problem.  As such,
this is a problem that
will probably have to be solved by some approximation scheme,
following such a solution
to the problem of the observables.

\subsection{Completeness of the physical state space}

Where do the exact physical states that have been found fit
into the whole space
of solutions?
This is a problem that is clearly dependent on the inner
product and physical observable
algebra; what we need in the end is to show that the physical
states carry a  representation of the physical observable algebra.

\subsection{Coupling matter to gravity}

The Ashtekar formalism allows coupling to all types of matter,
including
spin zero and one-half matter, Yang-Mills fields\cite{ART},
supergravity\cite{superted} and antisymmetric
tensor gauge theories\cite{lee-matter}.  It is easy to
extend the loop
representation to describe coupling to these
matter fields at the kinematical and diffeomorphism invariant level.
Nothing is known about solutions to the Hamiltonian
constraint including matter.

\section{Conclusions}

The results that I have been describing constitute a
collective work in progress, which has been  undertaken
by a number of people who share a common interest in the
questions I outlined in the introduction.  As with any result
of a scientific endeavor,
from Stonehenge down through the Macintosh computer on which I'm
writing this,
these results reflect both the knowledge and the aspirations of
those who made them.   While such a work remains unfinished, it
is difficult to judge its ultimate
worth.  We certainly don't yet know whether there is a consistent
quantum field theory that would go by the name of  general
relativity, although the steady progress we have been making
keeps us confident that it will be possible to cleanly resolve this
question.  However, this was, and is, not the only goal of
this program; it was equally hoped that this work would uncover
some general features that would hold for any quantum theory of
gravity that could be constructed nonperturbatively.  I believe
that
it is
fair to say that a number of such features have emerged, and
that as a result of this work we are wiser about how the world
will look when we have a satisfactory quantum theory of gravity
then we were before.  I would like to close by listing several
morals that I believe we have learned from the work I've described here.

a)  To solve the spatial diffeomorphism constraints it is
necessary to take a different starting point already at the
level of the quantum kinematics than is taken in conventional
Minkowski space quantum field theories.  To avoid introducing
background
structures, Fock space must be replaced by representations of
the kinematical observable algebra that rely on no background
metric and carry unbroken representations of the diffeomorphism
group.  That is, to get the
diffeomorphism invariant physics right, we must make sure that
our state space and regularization procedures are background
independent already at the kinematical level.

b)  At present the only representations known to have these
properties are the discrete representations I discussed here.
Whether or not there are others is presently an open question,
however even without resolving this, these new representations have
interesting structures that deserve more investigation.  In
essence, what they seem to do is to resolve the paradoxes that
follow from the uncountable nature of the classical continuum, as
each state in these representation has support only on a
countable set of loops.  It is exactly this structure that makes
it possible to solve directly the diffeomorphism constraints, in
a way in which the resulting space of diffeomorphism invariant
states has a countable basis.

The price we pay for this is that at the kinematical level the
state spaces are nonseperable.  This would be a serious problem
at the level of the physical state space; however it is only a
technical inconvenience in our case.

c)  It is a further property of this discrete
representation that it allows us to construct
finite and regularization independent operators
to represent non-local functions of the gravitational
field. This results in the quantization of the spectra of
areas and volumes.  This is, moreover, not a spurious
result of the kinematics, for we can show by direct
construction that the quantization of areas and volumes
is maintained at the diffeomorphism invariant level, when
they are measured by
diffeomorphism invariant operators.

We believe that these results will survive further
translation to the physical level.  If this is the case
they will be the first physical predictions made by a quantum
theory of gravity.  That is, we propose that any fine enough
measurement will reveal
that the area of any surface can only lie in a discrete spectrum
consisting of integral multiples of $l_{Planck}^2 /2 $, and certain
other values associated with intersections that are described
in \cite{review-lee}.

d)  The necessary dependence of renormalization procedures for
local operators on background metrics is, we believe, a
general phenomena.  As a result, I conjecture that in any quantum
theory of gravity there will be no renormalized local operators.
Further, all diffeomorphism invariant operators will be finite
after an appropriate {\it regularization}
procedure\cite{review-lee,carloun}.

e)  I believe that another thing we have seen in our
construction of a diffeomorphism invariant operator
in section 3.13 is quite general.  This is that all
diffeomorphism invariant operators, and hence all physical
operators, will measure topological
properties of non-local structures.

f)  This means that in the final quantum theory of gravity
we will see the continuous geometry of the classical
theory emerge from a quantum theory of purely topological
structures at the Planck scale.   This is a consequence of
what we discovered in
section 3.10 and 3.11, in which we saw that when the classical
limit was formulated carefully, it follows that every state
that behaves semiclassically at large scales must be far from
the semiclassical limit when probed on Planck scales.  So far, in
fact, that what is revealed is the  discrete structure
required by the quantization of the area operator.

g)  We believe that it is this behavior, which is apparent
already at the kinematical level, and not a pathology of
the dynamics of general relativity, that is responsible for
the failure of perturbative quantizations of general
relativity.  That is, the
perturbation theory is already wrong at the kinematical level
because it is unitarily inequivalent to the correct
kinematical state space.  Moreover, while at large distances
the correct physics can be well approximated by semiclassical
states, this
approximation becomes worse and worse at shorter and
shorter distances.

h) Finally, while the possibility of solving the diffeomorphism
constraint exactly is implied only by the existence of the loop
representation, which implies only that it is possible to choose
a connection as the canonical coordinate of the theory, that
it is in exactly the same representation that the
Hamiltonian constraint becomes exactly solvable seems the
main miracle uncovered so far.  (Here, by a miracle I mean
something wonderful that happens for a reason we don't
understand.)  It seems to be the
case that once the diffeomorphisms have been taken care of
correctly, the information remaining in the Hamiltonian
constraint is very manageable.

 i) Although I do not claim to understand completely what is
 behind this miracle,  it is worth pointing out that it, is
 in fact, exactly the existence of the self-dual connection
 that makes it possible to write the Hamiltonian constraint
 as a single
term, which in turn makes possible the exact solutions
which have been discovered.  It seems, as a result, very
possible that self-duality is one of the keys to quantum
gravity in the real $3+1$ dimensional world.    Indeed,
self-duality is the key to several very interesting results
that have been recently
uncovered about the classical theory \cite{CDJ,tedcarlo}.

 Note that only the last two of the nine morals in this list
 depended on the form of the Hamiltonian constraint, and hence
 on the
conjecture that general relativity is the correct microscopic
description of gravitation.  The rest depend only on the
existence of the loop representation, which needs only that the
theory can be expressed in such a way that a connection is the
canonical coordinate.  Thus, the fact that so many of the key
features are present at the kinematic and diffeomorphism invariant
levels, before the dynamics has been imposed, makes it, in
my opinion, quite likely that whatever dynamics turns
out to be
right, the description of Planck scale physics in the final
theory of quantum gravity will look a great deal like the
picture I have been sketching here.

\section*{Acknowledgements}

This contribution was intended as a report of work done by a
number of people and I am grateful to all of them for many
discussions about nonperturbative quantum gravity.  I am
particularly indebted to my collaborators in this work:
Abhay Ashtekar, Ted
Jacobson and
Carlo Rovelli;
in many places here I am expressing ideas that I
heard first from one of them.  Conversations
with Louis Crane over the years of the development
of this program have been important in
sharpening my understanding of what we are doing as
well as for killing any number of bad ideas.  I would
also like to thank John Baez, Julian Barbour,
Berndt Bruegmann, Riccardo Capovilla, John Dell,
Rudolfo Gambini, Joshua Goldberg, Gary Horowitz,
Viqar Husain, Karel Kuchar, Chris Isham, Lou Kauffman,
Roger Penrose, Jorge Pullin, Paul Renteln, Rafael Sorkin,
Madhavan Varadarajan and Edward Witten for
critical discussions about this work.  This work was
supported, in part,  by the National Science Foundation under grants
from the division of Gravitational Physics
and a US-Italy cooperative research program grant.

\end{document}